\documentclass[grl]{agu2001}
\usepackage{times}
\usepackage{amsmath}
\usepackage{amssymb}
\usepackage{graphicx}
\usepackage{psfrag}
\usepackage{setspace}
\def\url#1{\textcolor{blue}{\protect\small\sf #1}}
\usepackage{color}

\authorrunninghead{AGARWAL ET AL.}

\titlerunninghead{SEA ICE ALBEDO}

\authoraddr{J.~S. Wettlaufer, Yale University, New Haven, CT 06520-8109, USA.(john.wettlaufer@yale.edu)}

\begin{document}

%
%

\title{Decadal to seasonal variability of Arctic sea ice albedo}

%
%


\authors{S. Agarwal, \altaffilmark{1,2} W. Moon \altaffilmark{2} 
and J. S. Wettlaufer, \altaffilmark{2,3,4}}

\altaffiltext{1}{Department of Mathematics, Indian Institute of Technology Guwahati,
Guwahati 781 039, Assam, India}

\altaffiltext{2}{Department of Geology and Geophysics,
Yale University, New Haven, Connecticut, USA.}

\altaffiltext{3}{Department of Physics \& Program in Applied Mathematics, Yale University, New Haven, CT, 06520-8109, USA.}

\altaffiltext{4}{NORDITA,  Roslagstullsbacken 23, SE-10691 Stockholm, Sweden.}

%
%

\begin{abstract}
A controlling factor in the seasonal and climatological evolution of the sea ice cover is its albedo $\alpha$.  Here we analyze Arctic data from the Advanced Very High Resolution Radiometer (AVHRR) Polar Pathfinder and assess the seasonality and variability of broadband albedo from a 23 year daily record.  We produce a histogram of daily albedo over ice covered regions in which the principal albedo transitions are seen; high albedo in late winter and spring, the onset of snow melt and melt pond formation in the summer, and fall freeze up.  The bimodal late summer distribution demonstrates the combination of the poleward progression of the onset of melt with the coexistence of perennial bare ice with melt ponds and open water, which then merge to a broad peak at $\alpha \gtrsim $ 0.5.  We find  the interannual variability  to be dominated by the low end of the $\alpha$ distribution, highlighting the controlling influence of the ice thickness distribution and large-scale ice edge dynamics.  The statistics obtained provide a simple framework for model studies of albedo parameterizations and sensitivities.
\end{abstract}

%
%

\begin{article}

\section{Introduction\label{sec:intro}}

The rapid decline of Arctic sea ice coverage  during the last thirty years has led to substantial discussion and debate regarding the direct and indirect roles of the ice-albedo feedback, and other processes, in driving the potential loss of a perennial sea ice state \cite[e.g.,][]{OneWatt, SerrezeBarry:2011}.  The focus on ice-albedo feedback is based on intuition, modeling and observation \cite[e.g.,][and refs therein]{Walsh:2000}.  We understand that there are many mechanisms, operating over a wide range of length and time scales, that influence environmental change in the Arctic \cite[e.g.,][]{MLT:2009, SerrezeBarry:2011}.  Moreover, even in simple models the underlying operation of the direct and indirect ice-albedo feedback is a complex process \cite[e.g.,][]{Curry:1995, EW09, Renate:2011} and this has implications for Global Climate Models (GCMs) which  depend sensitively on the treatment of ice albedo  \cite[][]{EUW:2007, EUW:2008, DeWeaver:2008}.  Indeed, the fact that the IPCC AR4 GCMs underproject the decline of the ice during the satellite era \cite[e.g.,][]{OneWatt, SerrezeBarry:2011, Ian:2011} focuses attention on the methodology used to treat albedo in future model projections.   Field based measurements are essential for building physically based models of the processes that influence surface broadband albedo \cite[e.g.,][]{Perovich:2002, Skyllingstad:2009}, as well as for ground truthing of satellite retrieval algorithms \cite[e.g.,][]{Stroeve:2001}.   However, sparse coverage of in-situ measurements highlights the importance of combining field and space based methods to examine the spatiotemporal trends in sea ice albedo that provide an essential test bed for models \cite[][]{Perovich:2007}.  

\begin{figure}
\centering
\includegraphics[width=17pc]{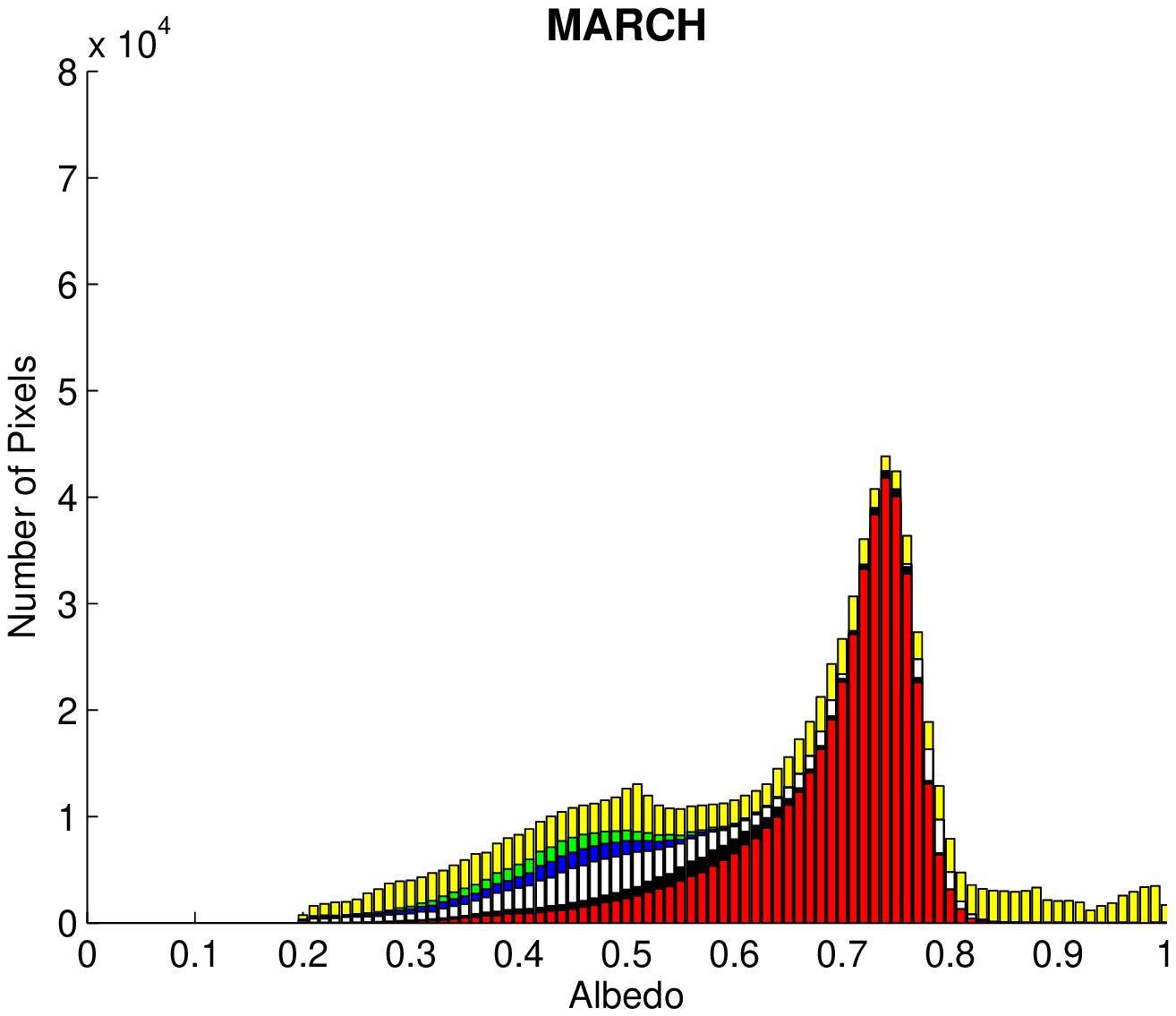}
\includegraphics[width=17pc]{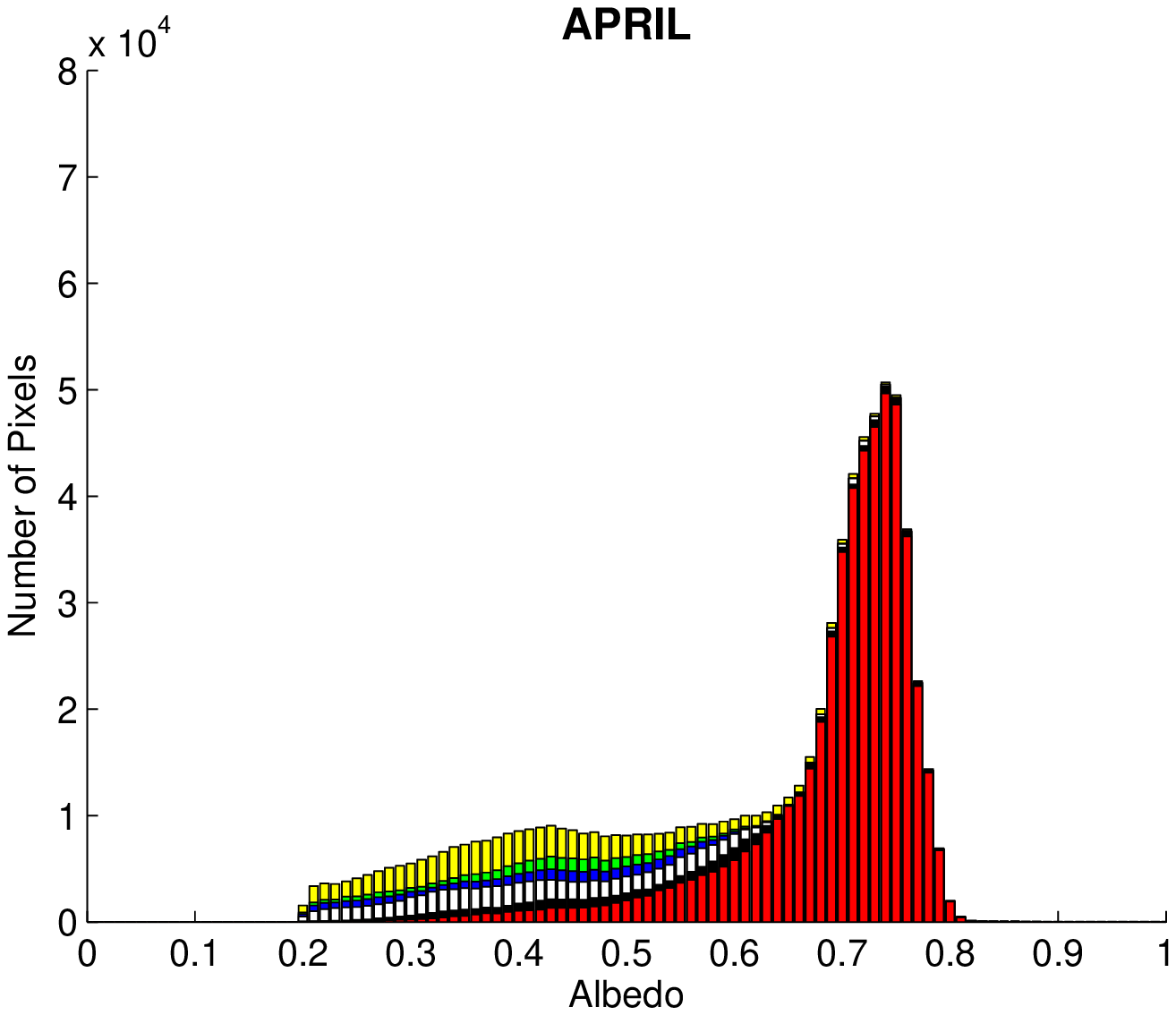}
\includegraphics[width=17pc]{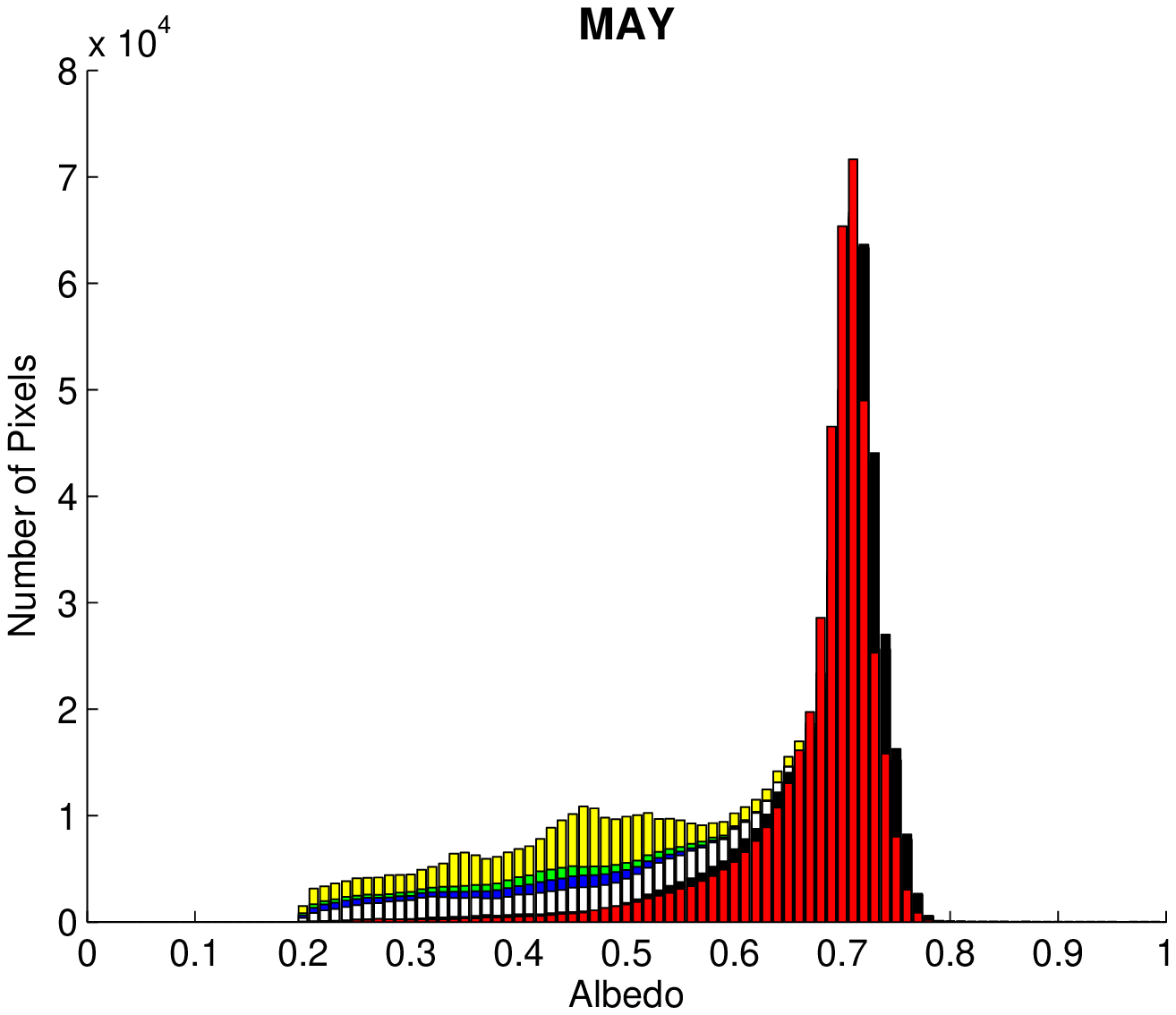}
\caption{The albedo histograms shown for days in mid-March, -April and -May.  If there is ice in a pixel for at least time $\tau_\text{th}$, then we compute the albedo for that pixel and average over all pixels that have met this criterion.  $\tau_\text{th}$ = 23 (Red), 22 (Black), 17 (White), 12 (Blue), 8 (Green) and 1 (Yellow) yr.  
}
\label{fig1}
\end{figure}

\begin{figure}
\centering
\includegraphics[width=17pc]{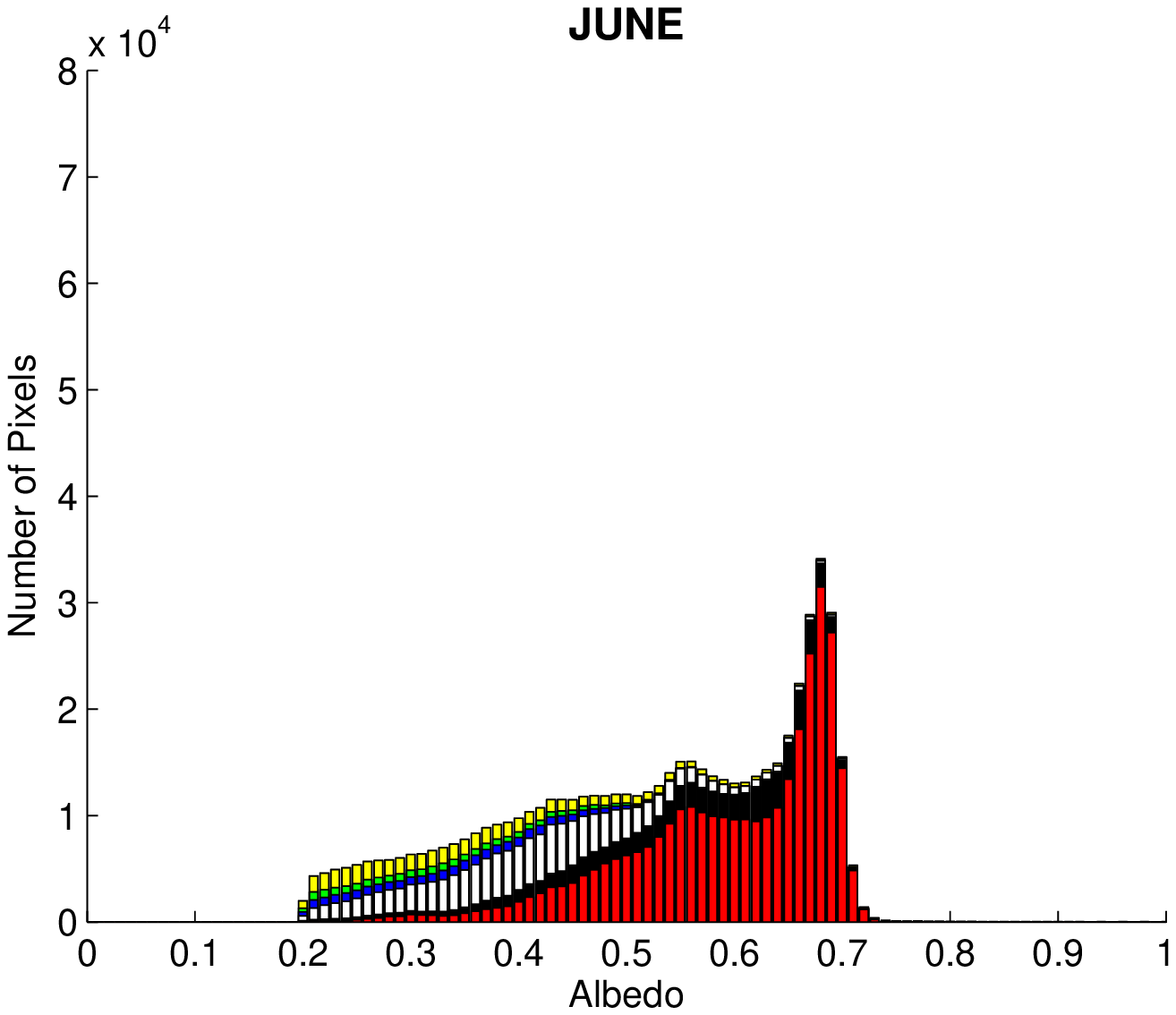}
\includegraphics[width=17pc]{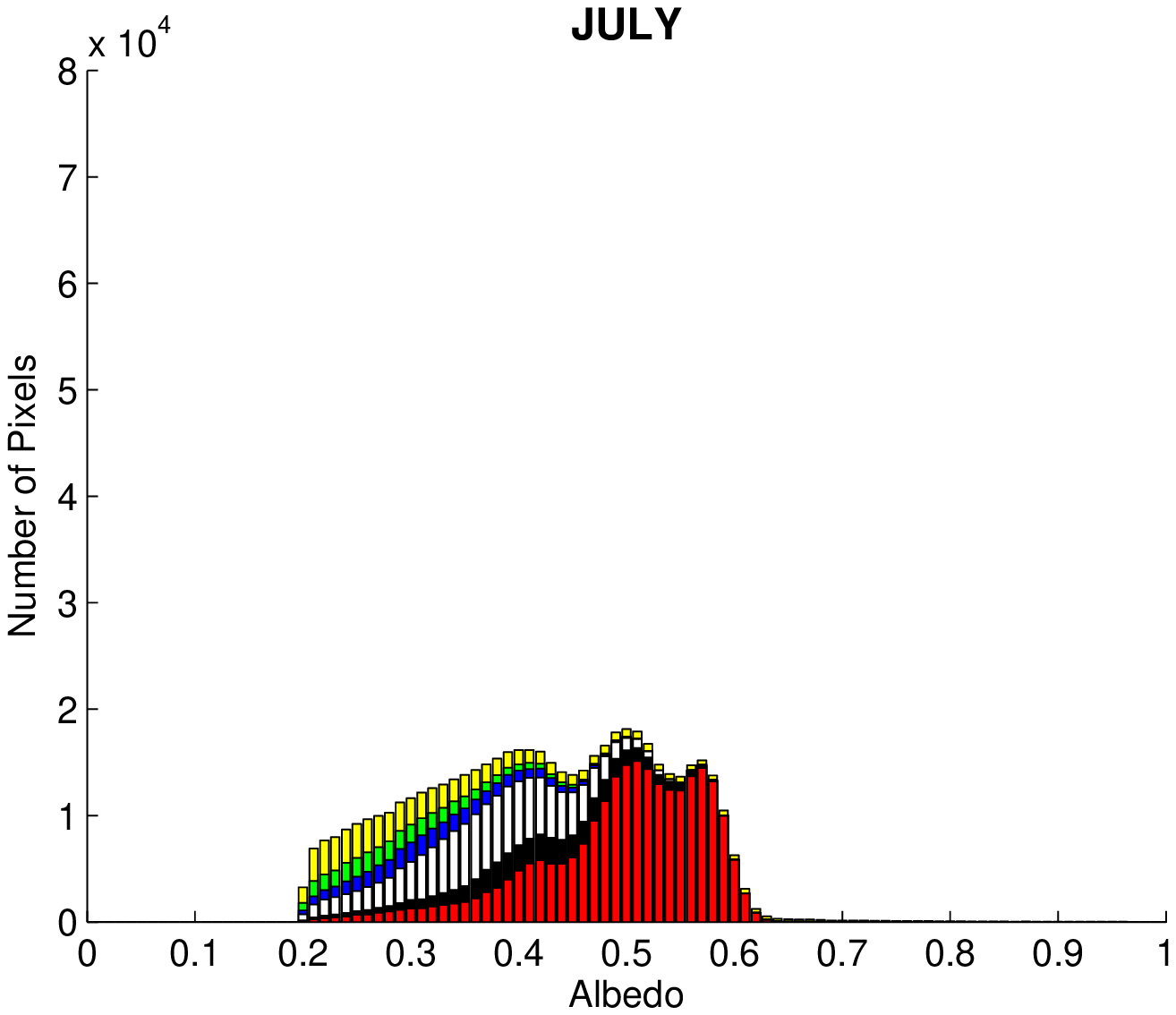}
\includegraphics[width=17pc]{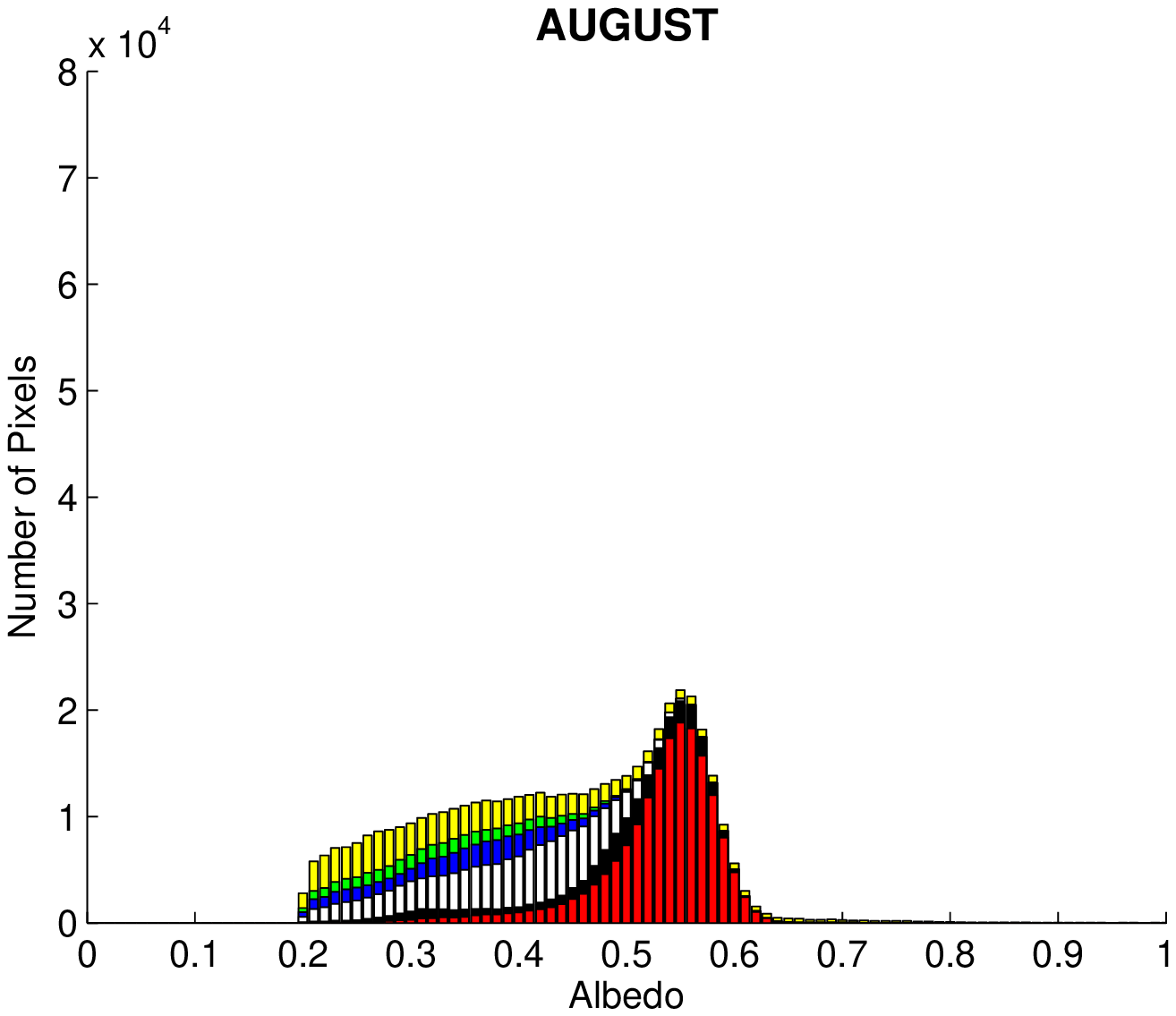}
\caption{The albedo histograms for mid-June, -July and -August.  If there is ice in a pixel for at least time $\tau_\text{th}$, then we compute the albedo for that pixel and average over all pixels that have met this criterion.  $\tau_\text{th}$ = 23 (Red), 22 (Black), 17 (White), 12 (Blue), 8 (Green) and 1 (Yellow) yr.  
}
\label{fig2}
\end{figure}

\begin{figure}[h!]
\centering
\includegraphics[width=17pc]{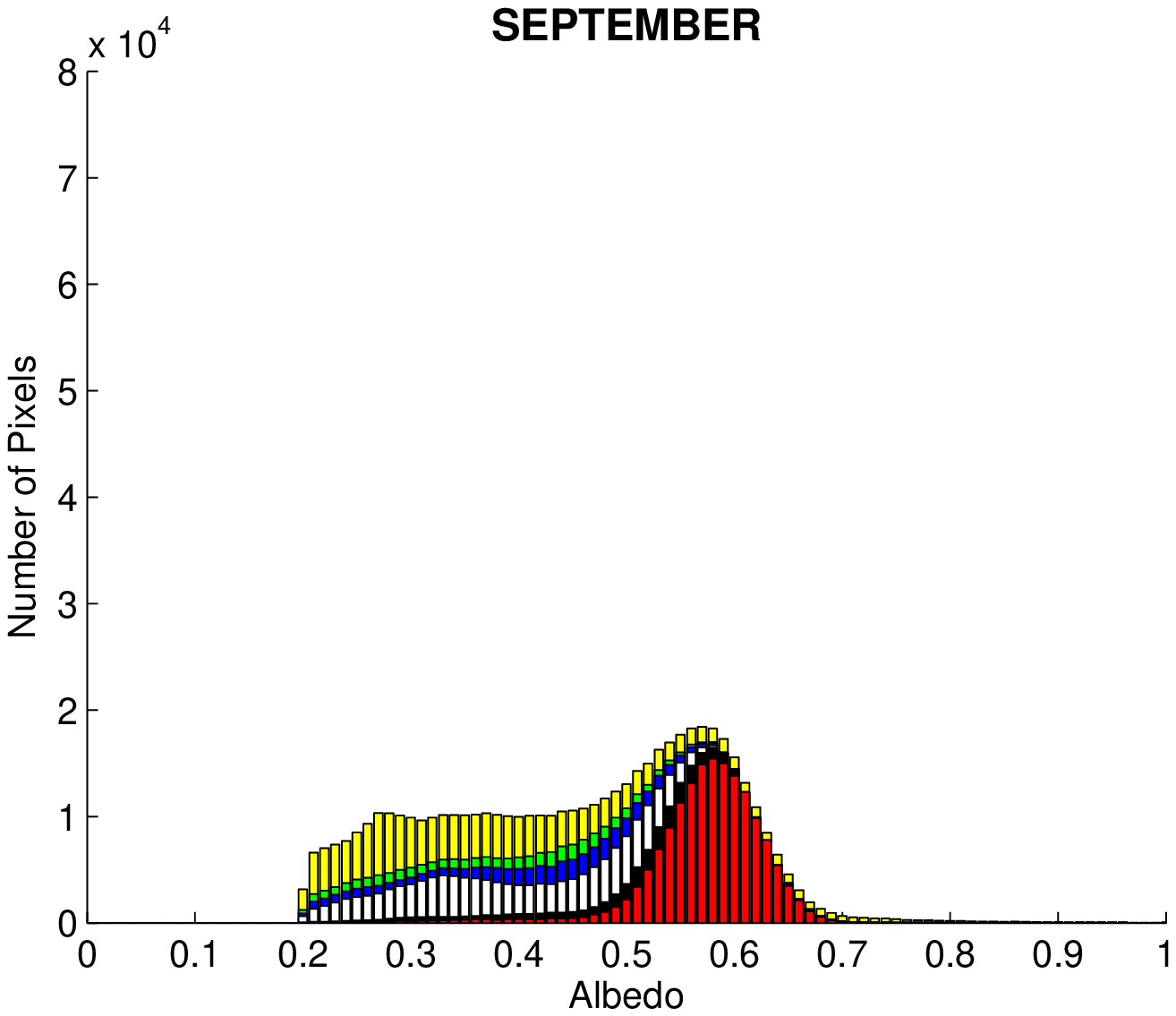}
\includegraphics[width=17pc]{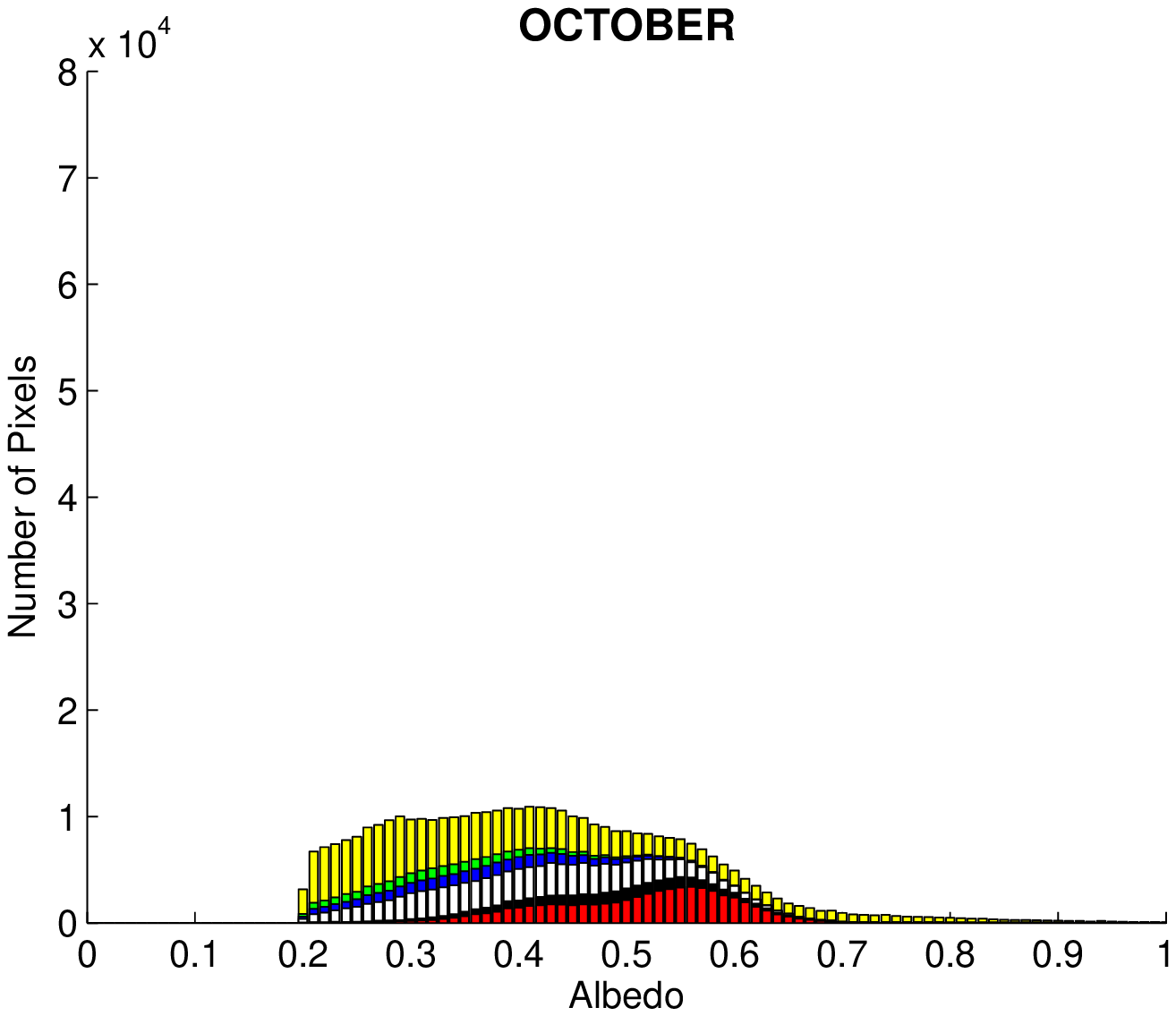}
\caption{The albedo histograms for mid-September and mid-October.  If there is ice in a pixel for at least time $\tau_\text{th}$, then we compute the albedo for that pixel and average over all pixels that have met this criterion.  $\tau_\text{th}$ = 23 (Red), 22 (Black), 17 (White), 12 (Blue), 8 (Green) and 1 (Yellow) yr.  The entire seasonal cycle is archived as a movie.   
}
\label{fig3}
\end{figure}

Here we describe an analysis of daily satellite retrievals of the directional - hemispheric apparent albedo from the Advanced Very High Resolution Radiometer (AVHRR) Polar Pathfinder (APP).  (The apparent albedo is what would be measured by upward and downward looking radiometers and hence varies with the state of the atmosphere and the solar zenith angle.  For simplicity we refer to this as $\alpha$.)  By culling out the albedo retrieved solely from the ice covered region of the Arctic Ocean from 1982 to 2004 a pixel scale view of the surface is created from which we produce a daily histogram.   
The principal seasonal surface transitions in the ice cover are observed; the high albedo snow covered spring to snow melt convolved with the poleward progression of melt onset, the appearance of melt ponds, their coarsening and fall freeze up (Figures \ref{fig1}-\ref{fig3}).  The data reveal the decadal and intra- and inter-seasonal variability of the albedo over multiple time scales.  It is hoped that their statistics and trends will be of use in the development of models ranging from the very simple to GCMs.

\section{Methods\label{sec:data}}

The APP dataset has been refined and applied for use in a wide range of polar studies and is described in detail on the NSIDC website \cite[][]{NSIDC}.  
AVHRR channels 1 - 5 range from the visible to the thermal infrared (0.58 - 12.5 $\mu$m) and measure Top of the Atmosphere (TOA) reflectances (1, 2, 3A) and brightness temperatures (3, 3B, 4, 5).  Additional information includes the solar zenith, relative azimuth and satellite elevation angles, cloud and orbit masks and universal time \cite[see Figures 3 and 4 of][]{NSIDC}. For the analysis reported here we use the so-called Surface Type Mask and  $\alpha$ retrievals from 1 January 1982 through 31 December 2004, taken daily at 1400 hours.  The albedo retrieval  involves four steps developed and tested by \citet[][]{Csiszar:1999}: (a) normalize  channels 1 and 2 with respect to solar zenith angle, (b) convert channels 1 and 2 narrow band reflectance to a TOA broadband reflectance, (c) correct the TOA broadband reflectance for anisotropy, and (d) convert the TOA broadband reflectance to a clear sky surface broadband albedo.  

The Northern Hemisphere is gridded in 1805 $\times$ 1805 pixels, with each pixel representing a 5~km $\times$ 5~km region.  Sea ice is distinguished from land and open water using SSM/I brightness temperatures and filtering the Surface Type Mask data with the NASA Team Sea Ice Algorithm which distinguishes between first-year (FYI) and multi-year  ice (MYI) concentrations.  The approach ascribes a MYI flag to a region containing at least 50 percent of this ice type.  We note that uncertainties depend on the season (e.g., melt pond fraction) and region (near the ice edge), along with the surface type categories \cite[][]{Comiso:1996}.  In particular, as will become relevant in our interpretation, errors are enhanced in the presence of surface ablation (especially during snow melt) because a small amount of liquid water alters the emission characteristics of the surface and it is difficult to distinguish between open water at the freezing point and wet ice surfaces.  In their ground truthing study of this retreival scheme \citet[][]{Stroeve:2001} note that the satellite albedo and that  measured by radiometers at the surface are equivalent and are functions of atmospheric conditions, solar zenith angle and the satellite elevation angle.  Finally, the cloud masking scheme was ground truthed by \citet[][]{Maslanik:2001} during the most difficult part of the cloud seasonal cycle (April-July) over boxes from 15 to 305 km on a side.  The maximum standard deviation of albedo was found to be 0.05 which reduced precipitously as the length scale decreased.  

On physical grounds the albedo data are filtered to remove any values greater that 1 or less than 0.2. We examine each pixel every day for the presence of ice.  
Then, for a specified time threshold $\tau_\text{th}$, we average $\alpha$ for that pixel. Different thresholds $\tau_\text{th}$ specify the minimum time for which a pixel contained ice. A histogram for each day of the year is thereby produced and represents the number of pixels for each average
albedo value bin for that day.
Different colors on the histogram specify different thresholds.  Red indicates that a pixel has contained ice for the whole time period under consideration, i.e. $\tau_\text{th}$ = 23 yr (1982 through 2004). Black specifies if a pixel has been ice for at least $\tau_\text{th}$ = 22 yr in the entire time period, which helps us to remove any bias introduced by any minimum during a year.  White denotes $\tau_\text{th}$ = 17 yr; blue $\tau_\text{th}$ = 12 yr; green $\tau_\text{th}$ = 8 yr; and yellow $\tau_\text{th}$ = 1 year in the entire time period.   Our approach is motivated by several factors.  (a) The knowledge of the great sensitivity of GCMs to albedo parameterizations \cite[][]{EUW:2007}.  (b) Our continued development of  theoretical treatments for large scale transitions in the ice cover that test the robustness of bifurcations in the state of the system to the ice-albedo (and other) feedback \cite[][]{TimeScales}.  (c) Extending these theories to understand the role of noise.  Whence, an observational understanding of the temporal variability, from month to month and interannually is essential, and is simply captured using the threshold idea.

\section{Results and Discussion\label{sec:results}}

The results are shown in Figures 1-3 and we discuss first the entire record shown in red ($\tau_\text{th}$ = 23 yr).   Because of the limited spatial resolution of the retrieval and the uncertainties in ice type detection and albedo estimation, we do not expect
that the numerical values averaged over the ice-covered regions as determined here to map directly onto field observations, but our results nonetheless nicely display the five distinct phases of albedo evolution from spring through fall: dry snow, melting snow, pond formation, pond evolution, and fall freeze-up \cite[see Figure 3 of][and references therein]{Perovich:2007b}.   The dry snow value seen here ($\sim$ 0.75) will by definition be lower than a field observation on MYI ($\sim$ 0.85) because our pixels contain both MYI and FYI , which have varying amounts of snow cover \cite[][]{OneWatt}, and near the ice edge the pixels include open water.
For March, April and May $\alpha$ has a single strong peak about a value indicative of snow covered ice, with an asymmetry that shows a sharp decay at the high values and a broader decay at low values capturing the range of ice types and snow covers that are expected over the basin in the spring.  
By the middle of June the distribution is clearly bimodal, with a high peak at about the melting snow covered MYI value ($\sim$ 0.70) and melt pond ($\sim$ 0.50) signatures beginning to come through as the peaks broaden and move to lower values by mid-July.  This is due to the poleward progression of the onset of melting that, in mid-June, can have gradients that span 100's of km taking a week to progress northward from 73 to 82 $^{\text{O}}$N \cite[][]{Winebrenner:1994}.  Such a large scale progression is clearly resolved by our method and hence the low latitude semi-zonal region of low $\alpha$ contributes to left hand peak in our distribution.
They finally merge to a single broad peak at a value indicative of fractional coverage of melt ponds and perennial ice \cite[see Figures 11, 13 and 14 of ][]{Skyllingstad:2009}.  By mid-September, when freeze-up is in full swing, the peak shifts slightly to the right centered around a perennial ice value of $\alpha\sim$ 0.6.  Finally insolation diminishes in October.  

\begin{figure}[h]
\centering
\includegraphics[width=21pc]{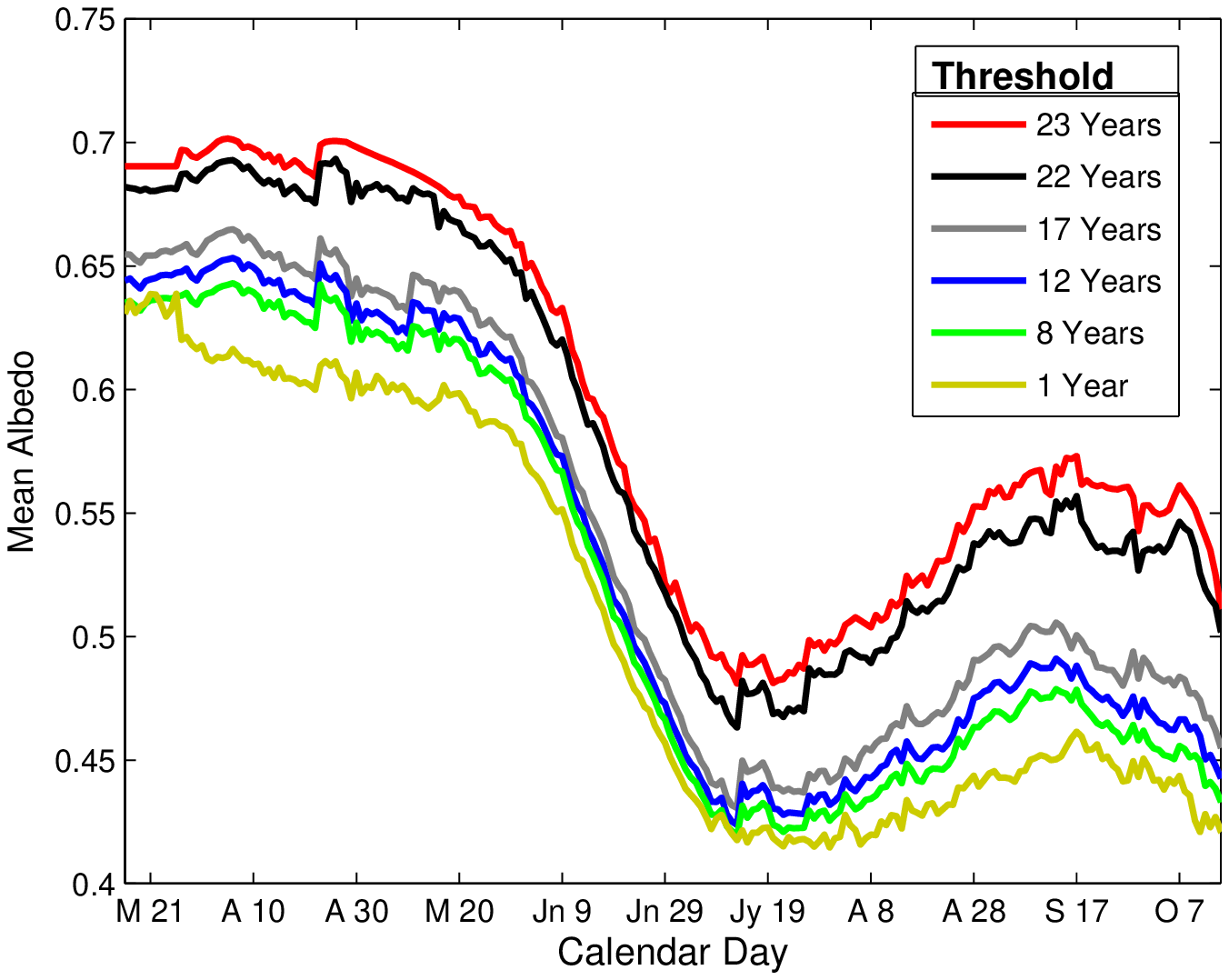}
\caption{The mean albedo covering the range of the previous figures, from mid March to mid October for the same thresholds; $\tau_\text{th}$ = 23 (Red), 22 (Black), 17 (Gray), 12 (Blue), 8 (Green) and 1 (Yellow) yr.  
}
\label{fig4}
\end{figure}

There is a systematic trend of increasing variability as $\tau_\text{th}$ decreases and this is dominated by the low $\alpha$ part of the distribution.  The variability increases substantially from decadal to seasonal, indicating both larger intrinsic fluctuations in the areal fraction of ice types as well as in the ice edge position.  Changes in both will have a signature at small values of $\alpha$.  The integrated influence of this observation is a larger interannual variability in solar insolation received by the Arctic Ocean, consistent with the findings of \citet[][]{Perovich:2007b}.  


As a means of further quantifying the variability, we plot the daily mean albedo as a function of $\tau_\text{th}$ in Figure \ref{fig4} which demonstrates a monotonic increase in the albedo with $\tau_\text{th}$.  The seasonal trend stands out, as does the decrease in the threshold dependence during the early evolution of the melt season.  The latter observation is consistent with the notion that as snow and ice melt from the surface and the system approaches the bulk freezing point, the intrinsic variability between ice types and ice and water is suppressed.  Moreover,  as discussed above, because the retrieval scheme is less accurate as the pack begins to ablate from its surface \cite[][]{Comiso:1996}, there is less $\tau_\text{th}$ dependence during these periods.  However, as the open water, melt ponds and mature ice surfaces in the pack evolve toward freeze up, curves with different $\tau_\text{th}$ begin to diverge as might be expected \cite[][]{Skyllingstad:2009, Fetterer:1998}.  Finally, in Table S1 we give the monthly variance in the albedo as a function of $\tau_\text{th}$ which provides a slightly different view of the seasonal evolution shown in Figure \ref{fig4}.

The annual cycle of low stratiform cloud fraction ranges from 20\% in the winter to 70\% in the summer.  According to the observational synthesis of \citet[][]{Eastman:2010} the interannual variations of cloud amounts demonstrate significant correlations with surface air temperature, and total sea ice area as well as the Arctic Oscillation.  Septembers with low areal extent are generally preceded by a summer with decreased middle and precipitating clouds and the autumns following such Septembers are enhanced in low cloud cover. They conclude that the total cloud cover appears to be greater throughout the year during low-ice years.  Hence, as expected, this increasing cloud cover would promote long wave forcing and ice loss, which may also be enhanced by the observed decrease of summer cloud cover.  Interestingly, when the cloud fraction is growing from April to mid June (20-70\%) viz., the climatological seasonal cycle, we see the least dependence on threshold (Figure \ref{fig4}) which we believe is due to the suppression of the difference in ice type and water associated with the ablation of the surface.  Therefore, an observed trend of decreasing cloud cover in summer would be consistent with the weaker threshold dependence we observe for the onset of melt and the increase in the threshold dependence during autumn freeze up, because of the larger fraction of open water.  However, any increase in cloud fraction correlated with minima in ice cover would manifest itself as an increase in the low-$\alpha$ part of the distribution with decreasing $\tau_\text{th}$ as already seen here.  Thus, to ascribe a cause would require independent measurements, on the same time scales, of all cloud types.  Interpretation would also benefit from a clear distinction between the albedos of the different ice types which is the topic of a more extensive study. 

\section{Conclusions}

We have analyzed daily retrievals of the directional - hemispheric apparent albedo from summer 1982 to summer 2004 using the APP archive and have produced histograms of daily albedo over ice covered regions.  This reveals the principal albedo transitions over the ice pack; high albedo associated with dry snow-covered ice in late winter and spring, the onset of snow melt and melt pond formation and evolution in the summer, and fall freeze up. We find a bimodality of the late summer histogram, displaying the combination of the poleward progression of the onset of melt with the  coexistence of perennial bare ice with melt ponds which eventually merge to a broad peak at $\alpha \gtrsim $ 0.5 consistent with field constrained models \cite[e.g.,][]{Skyllingstad:2009}.  There is substantial decadal to interannual variability which is dominated by the low end of the $\alpha$ distribution.  This is consistent with a recent combined field and remote sensing study that found large interannual variability in the solar insolation in the Arctic \cite[][]{Perovich:2007b}, and our method allows us to examine variability on multiple time scales up to the length of the record. In the simple approach described here the variability originates in the natural fluctuations associated with the interannual changes in the fractions of ice types and in the position of the ice edge.  By implication, a similar study of the more recent past would presumably reveal a broad low albedo distribution associated with the increased fraction of FYI \cite[see Figure 2 of][]{OneWatt}.  Because of the thickness dependence of the albedo \cite[see e.g., Eq. 4 of][]{EW09} and the observed increasing fraction of thin ice, the results highlight the role of the ice thickness distribution in controlling the spatially homogenized albedo.  This has implications for the manner in which all classes of models, from the very simple to GCMs, treat this important quantity.  Such models could easily utilize our distributions or their moments as input and use the $\tau_\text{th}$-dependent variances as a baseline for treating the variability.  


%
%
%
%
%
%

%
%
%
%

\begin{acknowledgments}
The authors thank I. Eisenman, R. Kwok and N. Untersteiner for feedback on various aspects of this project as well as the three anonymous referees 
for their thorough comments. They also thank Yale University for support.  
WM thanks NASA for a graduate fellowship and JSW thanks the Wenner-Gren Foundation and the John Simon Guggenheim Foundation.  

\end{acknowledgments}



%
%
%
%
%
%
%
%
%
%

\end{article}
\end{document}